\documentclass[aps,prl,twocolumn,showpacs]{revtex4}

\usepackage{graphicx}
\def  \bsig    {\mbox{\boldmath$\sigma$}}

\def \ba       {\mbox{\boldmath$a$}}
\def \bO       {\mbox{\boldmath$\Omega$}}

\begin{document}

\title{Topological Hall effect and Berry phase in magnetic nanostructures}
\author{P. Bruno$^1$, V.~K.~Dugaev$^{1,2}$, and M. Taillefumier$^{1,3}$}
\affiliation{$^1$Max-Planck-Institut f\"ur Mikrostrukturphysik,
Weinberg 2, 06120 Halle, Germany\\
$^2$Institute for Problems of Materials Science, NASU,
Vilde 5, 58001 Chernovtsy, Ukraine\\
$^3$Laboratoire Louis N\'eel, CNRS, Boite Postale 166,
38042 Grenoble Cedex 09, France}
\date{\today }

\begin{abstract}
We discuss the anomalous Hall effect in a two-dimensional electron
gas subject to a spatially varying magnetization. This topological
Hall effect (THE) does not require any spin-orbit coupling, and
arises solely from Berry phase acquired by an electron moving in a
smoothly varying magnetization. We propose an experiment with a
structure containing 2D electrons or holes of diluted magnetic
semiconductor subject to the stray field of a lattice of magnetic
nanocylinders. The striking behavior predicted for such a system
(of which all relevant parameters are well known) allows to
observe unambiguously the THE and to distinguish it from other
mechanisms.

\vskip0.5cm \noindent
\end{abstract}
\pacs{73.20.Fz; 72.15.Rn; 72.10.Fk}
\maketitle

After half a century of theoretical efforts, the Hall effect of
ferromagnets (usually called {\it anomalous} or {\it
extraordinary} Hall effect) remains a puzzling and controversial
topic. Until recently, it was considered that it originates from
the combined effect of exchange and spin-orbit (SO) interactions.
Two mechanisms of anomalous Hall effect (AHE) have been identified
(the {\em skew scattering} \cite{karplus54,smit58} and {\em
side-jump} \cite{berger70}) and studied thoroughly \cite{hurd}.
Recently, a new point of view has been proposed \cite{m_onoda02},
in which the AHE is expressed in terms of a Berry curvature in
momentum space. However, one should note that a generally accepted
theory treating on an equal footing all the above mentioned
contributions to the SO-induced AHE is still missing.

Recently, it has been suggested that (in addition to the above
mentioned SO-induced mechanism) a new mechanism may give rise to a
non-vanishing Hall effect in ferromagnets having a topologically
non-trivial (chiral) spin-texture, such as manganites or
pyrochlore-type compounds \cite{matl98}. To distinguish this
mechanism from the SO-based mechanism, we shall refer to it
hereafter as the {\em topological Hall effect} (THE).

Several theoretical papers have been devoted to the THE. In order
to explain the AHE observed in manganites, a model of 3D
ferromagnet with thermally excited skyrmion strings (topological
dipoles)has been proposed \cite{ye99}, showing that a THE can be
induced by the Berry phase \cite{berry84} related to the spatial
variation of magnetization in the vicinity of the string. The case
of disordered ferromagnets in the limit of small exchange
splitting has been addressed in Ref.~\onlinecite{tatara02}. In
both cases, in order to get a net topological field (or
chirality), the SO-coupling must be invoked. Even when the net
topological field vanishes, a nonvanishing THE may be obtained, as
discussed for a 2D kagom\'e lattice or a 3D pyrochlore lattice
\cite{nagaosa}.

All the above mentioned discussions of the THE concern systems
with spin-chirality at the microscopic scale (e.g., pyrochlore
lattice) or due to skyrmion-strings. In both cases, quantitative
experimental information on the chirality is not easily available.
Furthermore, the SO-mechanism is usually also present, which makes
complicate the quantitative interpretation of the observations.

In the present Letter, we propose to investigate the THE in
nanostructures, namely in a 2D electron (or hole) gas, in which a
well-controlled artificial chirality can be induced from the stray
field of a lattice of magnetic nanocylinders. The great advantage
of such a model system is that, in contrast to the above mentioned
cases, all relevant parameters are well known and (to some extend)
adjustable. We show that it is possible to obtain a significant
net topological field, and that the latter has a characteristic
variation with respect to a (uniform) external field, providing an
unambiguous signature of the THE. Finally, we suggest a system
appropriate for an experimental check of our theory.

We start from a model of 2DEG in a smoothly varying magnetization
${\bf M}({\bf r})$. The Hamiltonian has the following form
\begin{equation}
\label{1}
H=-\frac{\hbar ^2}{2m}\,
\frac{\partial ^2}{\partial {\bf r}^2}\;
-g\, \bsig \cdot {\bf M}({\bf r}),
\end{equation}
where $g$ is the coupling constant and $\bsig $ is the vector of
Pauli matrices. We assume that the amplitude of magnetization is
constant, ${\bf M}({\bf r})=M\, {\bf n}({\bf r})$, and that the 3D
unit vector ${\bf n}({\bf r})$ is a slowly varying function of
coordinates.

We use a gauge transformation $T({\bf r})$, which makes the
quantization axis oriented along vector ${\bf n}({\bf r})$ at each
point \cite{tatara97}. It transforms the last term in (1) as
$T^\dag ({\bf r})\, \left[ \bsig \cdot {\bf n}({\bf r})\right]
\,T({\bf r})=\sigma _z$, corresponding to a local rotation of the
quantization axis from $z$ axis to the axis along ${\bf n}({\bf
r})$.

The transformed Hamiltonian describes the electrons moving in a
(spinor) gauge potential ${\bf A}({\bf r})$,
\begin{equation}
\label{2}
H^\prime =-\frac{\hbar ^2}{2m}\,
\left( \frac{\partial }{\partial {\bf r}}
-\frac{ie}{\hbar c}\; {\bf A}({\bf r}) \right)
^2-gM\, \sigma _z\, ,
\end{equation}
where $A_i({\bf r}) =2\pi i\, \phi _0\; T^{\dag }({\bf r})\;
\partial _i\, T({\bf r})$, $\phi _0=hc/e$ is the flux quantum, and
$i=x,y$. For convenience, we define the gauge potential ${\bf
A}({\bf r})$ to have the same dimension as the electromagnetic
vector potential. The components of ${\bf A}({\bf r})$ can be
found easily using an explicit form of $T({\bf r})$.

Hamiltonian (\ref{2}) with the spinor ${\bf A}({\bf r})$ contains
terms inducing transitions between the spin-polarized states. We
consider the case when spin-flip transitions can be neglected,
i.e., where the spin adiabatically follows ${\bf n}({\bf r})$. It
corresponds to the condition $\lambda \equiv (\varepsilon
_F/\varepsilon _0)\, (k_F\xi )^{-1}\ll 1$, where $\xi $ is a
characteristic length of the variation of ${\bf n}({\bf r})$,
$\varepsilon _F$ is the Fermi energy, and $\varepsilon _0=2gM$ is
the spin splitting.

We also assume that the 2DEG is half-metallic with the Fermi level
located in the spin-up subband. Then we can neglect
the spin-down electrons, and we obtain the following effective
Hamiltonian for spinless electrons:
\begin{equation}
\label{3}
{\tilde H}=-\frac{\hbar ^2}{2m}\;
\left( \frac{\partial }{\partial {\bf r}}-\frac{ie}{\hbar c}\; \ba ({\bf r})
\right) ^2+V({\bf r}),
\nonumber
\end{equation}
where
\begin{equation}
\label{4}
a_i({\bf r})=
-\frac{\pi \phi _0\, \left( n_x\, \partial _in_y-n_y\, \partial _in_x\right)}
{1+n_z}\; ,
\end{equation}
the potential $V({\bf r})=(\hbar ^2/8m)\, (\partial _in_\mu)^2 $
results from the second order in ${\bf A}({\bf r})$ terms, and
$\mu =x,y,z$. For spin-down electrons the sign of the gauge field
$\ba ({\bf r})$ in Eq.~(\ref{3}) is reversed.

The topological field is defined as $B_t=\partial _xa_y-\partial
_ya_x$. It acts on the electrons within the spin polarized subband
like the ordinary magnetic field, and in particular gives rise to
a Lorentz-type force \cite{aharonov}. Using (4) we find
\begin{eqnarray}
\label{5}
B_t=-\frac{\phi _0}{4\pi }\; \epsilon _{\mu\nu\lambda}\,
n_\mu \, (\partial _xn_\nu )\, (\partial _yn_\lambda ),
\end{eqnarray}
where $\epsilon _{\mu\nu\lambda}$ is the unit antisymmetric tensor.
The integral over the area $S_0$ enclosed by an arbitrary contour $L_0$
\begin{equation}
\label{6}
\Omega (L)=\frac12 \int _{S_0} d^2{\bf r}\;
\epsilon _{\mu\nu\lambda}\, n_\mu \, (\partial _xn_\nu )\, (\partial _yn_\lambda )
\end{equation}
is the Berry phase calculated as the spherical angle spanned by an
area $S$ inside the contour $L$ in ${\bf n}$-space. This results
from the mapping of the contour $L_0$ onto the
contour $L$ in the mapping space $S_2$. In the 2D case with a
constant magnetization at infinity, we can compactify 2D plane to
a sphere $S_2$, and the quantity $I\equiv \Omega /4\pi $
calculated with the integral (6) over $S_2$ is the topological
invariant corresponding to the number of
covering the mapping space.

We can consider different topologically nontrivial types of the
distribution of magnetization field in 2D plane like, for example,
separated magnetic domains embedded into homogeneous
magnetization. Each domain creates a unit flux of topological
field $+\phi _0$ or $-\phi _0$, where the sign is related to the
magnetic polarization inside the domain with respect to the
ambient. The important point is that this topological flux does
not depend on a specific shape or a size of the domain. Besides,
it does not depend on whether the domain is separated by the
domain wall of Bloch or N\'eel type. The same result will be also
for the skyrmion - a topological excitation similar to the
circular magnetic domain with a N\'eel domain wall.

\begin{figure}
\begin{center}
\includegraphics[scale=0.3]{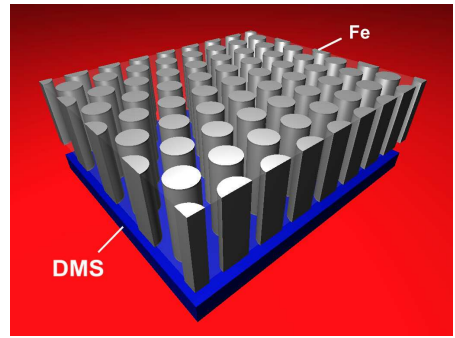}
\caption{The proposed structure consisting of a triangular lattice
of magnetic nanocylinders on top of 2D diluted magnetic
semiconductor (DMS).}
\end{center}
\end{figure}

Next we consider a periodic magnetization created by a regular 2D
lattice of the elements with a nontrivial topology. For example,
it can be a square or triangular lattice of circular domains. Each
of the domains creates the same magnetic flux. We can estimate the
THE for this system using the formula for the Hall conductivity in
the average topological field $\overline{B_t}$.
\begin{equation}
\label{7}
\sigma _{xy}({\bf r})
=\frac{2\pi \hbar e^2n_\uparrow\tau _\uparrow ^2\overline{B_t}}{m^2\phi _0},
\end{equation}
where $n_\uparrow $ and $\tau _\uparrow$ are the concentration and momentum
relaxation time of spin up electrons, respectively.
If the system is not half-metallic, there is also a corresponding
additional contribution to $\sigma _{xy}$, which differs from (7) by sign.

\begin{figure}
\begin{center}
\includegraphics[scale=0.3]{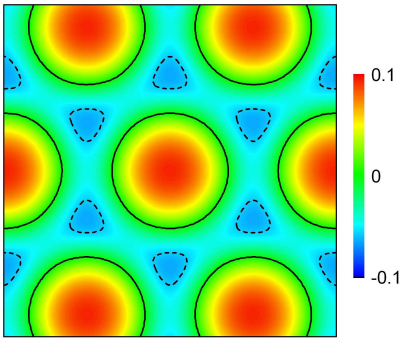}
\caption{Distribution of the $z$-component of dipolar field
$B/4\pi M_s$ inside the semiconductor film for the triangular
lattice of magnetic nanocylinders, for a zero external field. The
black solid circles correspond to the lines with $B_z=0$. Dashed
lines correspond to the lines with $B_z=0$ under an uniform
external magnetic field $B_{\rm ext}/4\pi M_s=+0.058$.}
\end{center}
\end{figure}

\begin{figure}
\begin{center}
\includegraphics[scale=0.3]{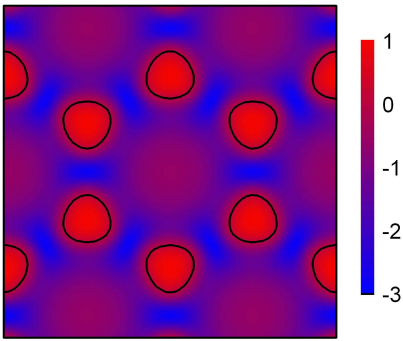}
\caption{Topological field $B_t({\bf r})$ (in units of $\phi _0$
per unit cell area) for the triangular lattice of magnetic
nanocylinders. Black lines correspond to $B_t=0$.}
\end{center}
\end{figure}

\begin{figure}
\begin{center}
\includegraphics[scale=0.3]{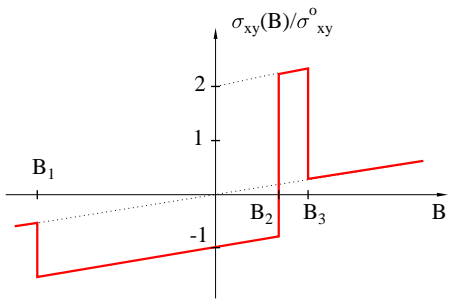}
\caption{Dependence of the Hall conductivity on external magnetic
field (schematically). The slope corresponds to the contribution
of the normal Hall effect; $\sigma_{xy}^0$ is the Hall
conductivity corresponding to a total (magnetic $+$ topological)
flux per unit cell equal to $\phi_0$.}
\end{center}
\end{figure}

Using (7) and the standard formula for conductivity, $\sigma
_{xx}=n_\uparrow e^2\tau _\uparrow/m$, in the case of unfilled
spin-up subband we find
\begin{equation}
\label{8}
\rho _{xy}\simeq \frac{\sigma _{xy}}{\sigma _{xx}^2}
=\frac{\overline{B_t}}{ecn_\uparrow }.
\end{equation}

The periodic magnetization can be realized using a structure with
a lattice of magnetic nanocylinders \cite{nielsch01} on top of 2D
electron gas. To observe the THE we suggest using II-VI diluted
magnetic semiconductor (DMS), as shown schematically in Fig.~1.
The stray field from the magnetic nanocylinders, penetrates into
semiconductor and polarizes the magnetic impurities, creating a
large spin splitting of electrons or holes in the DMS
\cite{furdyna88}. This effect is much stronger than the direct
diamagnetic or Zeeman actions of external magnetic field. The main
idea is to design a system, which provides a non-uniform magnetic
field of zero average within the semiconductor but
induces a topological field of nonzero average.

We have calculated the stray field and topological field for a
triangular lattice of nanocylinders (of magnetization $M_s$ along
the $z$-axis), using realistic values of parameters for the
nanocylinder lattice. Namely, we take a lattice constant of
100~nm, a cylinder radius of 37~nm (i.e., a filling ratio of 50
percent), and the gap $D$ between the lattice and semiconductor
equal to 20~nm. The distribution of $z$-component of the dipolar
field (normal to the semiconductor surface) is shown in Fig.~2.
The corresponding topological field calculated from Eq.~(5) is
presented in Fig.~3. For Fe cylinders, the dipolar field $B$
inside the semiconductor is about 2~kG in absolute value. The
characteristic length for the variations of ${\bf n}({\bf r})$ is
of the order of the gap $D$, i.e., $\xi\simeq 20$~nm.

One can see quite easily (for example by considering the Berry
phase from the mapping of the plane onto the sphere $S_2$) that
the net topological flux per unit cell, $\phi_t$, has to be an
integer multiple of $\phi_0$, and that $\phi_t/\phi_0$ is given by
the number (per unit cell) of lines $B_z=0$ enclosing a region
with $B_z<0$ minus the number of lines $B_z=0$ enclosing a region
with $B_z>0$. From Fig.~2, one thus sees that we have a triangular
lattice of lines $B_z=0$ enclosing regions with $B_z<0$, yielding
$\phi_t =-\phi_0$ for a vanishing external field. For the
considered geometry this corresponds to an average topological
field of about $-5$~kG, with local values ranging between $-15$~kG
and $+5$~kG.

If a negative external field is applied (assuming that the
magnetization of the nanocylinders remains unchanged, due to some
large coercivity), the lines with $B_z=0$ shrink without changing
their topology until they collapse to a single point and
eventually disappear at a critical field $B_1$, beyond which
$\phi_t=0$. If a positive field is applied, the lines with $B_z=0$
expand, until they connect each other at a critical field $B_2$
and change their topology to a honeycomb lattice (the dual of the
triangular lattice) of lines $B_z=0$ enclosing regions with
$B_z>0$ (dashed lines in Fig.~2), giving $\phi_t = +2\phi_0$.
Increasing further the external field leads to a collapse of the
lines with $B_z=0$ at a critical field $B_3$, beyond which
$\phi_t=0$. For Fe nanocylinders in the above geometry, the
critical fields are, respectively, $B_1\simeq -2$~kG, $B_2\simeq
+0.9$~kG, and $B_3\simeq +1.3$~kG. This change of topology under
application of an external magnetic field results in a striking
field dependence of the Hall resistivity, as sketched in Fig.~4,
and constitutes an unambiguous signature of the THE.

For the practical realization, we propose to use $p$-type DMS
since the exchange constants for holes are much larger than for
electrons \cite{furdyna88
}. For the estimation of the
topological field acting on holes, we use the Luttinger
Hamiltonian \cite{luttinger56}
\begin{equation}
\label{9}
H_h=\frac{\hbar ^2}{2m_0}\left[
\left( \gamma _1+\frac52\, \gamma _2\right) \nabla ^2
-2\gamma _2 ({\bf J}\cdot \nabla )^2\right]
+E_{ex}\, {\bf J}\cdot {\bf n}({\bf r}),
\end{equation}
where $J_\mu $ are the matrices of momentum $J=3/2$,
$\gamma _1$ and $\gamma _2$ are the Luttinger parameters, $m_0$ is
the free electron mass, ${\bf n}({\bf r})$ is the unit vector
along the direction of exchange field acting on 2D gas of holes,
$E_{ex}$ is the average exchange field created by
magnetic impurities,
\begin{equation}
\label{10}
E_{ex}=-J S x N_0\beta \,
\mathcal{B}_S(g_L\mu _BB/k_BT) ,
\end{equation}
$N_0\beta $ is the $p$-$d$ exchange constant, $S$ is the magnetic
moment of impurity, $xN_0$ is the concentration of magnetic atoms,
$\mathcal{B}_S(z)$ is the modified Brillouin function, and
$g_L$ is the Land\'e factor of magnetic atom.

We calculate the topological field for holes described by
Hamiltonian (9), using the transformation $T({\bf r})=\exp \, (i\,
{\bf J}\cdot \bO ({\bf r})/2)$, where $\bO ({\bf r})$ is the
vector of rotation. We find the topological field $B_t$
considering small deviations of the vector ${\bf n}({\bf r})$ in a
vicinity of some point ${\bf r}$. Assuming that the magnetic
splitting under exchange field is strong, we find that the gauge
potential and the topological field for holes in the valence band,
being proportional to $J$, are given by the same expressions
(\ref{4}) and (\ref{5}) as for electrons, multiplied by a factor
3.

After transforming (9) and restricting ourselves by the subband
$J_z=-3/2$, we obtain for spin-polarized holes, the same
Hamiltonian as (\ref{3}), with $-1/m$ replaced by $1/m^\star =
(\gamma_1 +\gamma_2 )/m_0$.

The spin splitting in II-VI semiconductor can be large enough to
provide 100\% polarization of holes under the dipolar field of
2~kG. Indeed, we can take the exchange coupling $N_0\beta
=-1.2$~eV, which is the typical magnitude for different compounds
(Cd$_{1-x}$Mn$_x$Se, Zn$_{1-x}$Mn$_x$Se, etc), and the atomic
density of magnetic atoms $x=0.05$. For $B=2$~kG, $g_L=2$, and
$T=4.2$~K the Brillouin function $\mathcal{B}_{5/2}(g\mu
_BB/k_BT)\simeq 0.075$. Then using (10) we obtain $E_{ex}\simeq
11$~meV. The Fermi energy of 2D holes filling the spin-splitted
subband is determined by the density of holes $n_p$, $E_F=2\pi
\hbar^2\, n_p/m^*$, where $m^*$ is the effective mass of holes. By
taking $m^*=0.5\, m_0$, and $n_p=10^{11}$~cm$^{-2}$, we obtain
$E_F\simeq 1$~meV. Thus, the condition of 100\% polarization of
holes, $E_F<E_{ex}$, is quite realistic for II-VI diluted magnetic
semiconductors. The adiabaticity parameter for this choice of
parameters is $\lambda \simeq 0.04\ll 1$. It should be noted that
much stronger splitting can be reached for Zn$_{1-x}$Cr$_x$Te
compounds, for which $N_0\beta =3.6$~eV \cite{mac96}.

For electrons, on the other hand, in order to fulfill the
conditions of full polarization and adiabaticity, lower
temperatures and/or very low electron densities would be needed.
Taking the exchange constant of $s$-$d$ interaction $N_0\alpha
=0.22$~eV \cite{furdyna88}, the electron effective mass
$m_e=0.22\, m_0$, $T=4.2$~K, and the concentration of electrons
$n_e=10^{11}$~cm$^{-2}$, we obtain $E_F\simeq 4.4$~meV,
$E_{ex}\simeq 2$~meV, and $\lambda \simeq 1$; the condition of
adiabaticity is poorly fulfilled, resulting in a reduction of the
effect. Also the gas is not fully polarized, reducing further the
THE due to the partial compensation of contributions from spin up
and down subbands. Although not fully developed, the THE should
nevertheless be observable with electrons.

In conclusion, we have proposed that the THE can be observed in
suitably chosen nanostructures, and that its striking behavior
under an external magnetic field provides an unambiguous
experimental signature of the THE.

\begin{acknowledgments}
We thank J. Cibert for discussions. V.D. thanks Polish State
Committee for Scientific Research for support under Grants
PBZ/KBN/044/P03/2001 and 2P03B05325. P.B. acknowledges financial
support from BMBF (Grant No.~01BM924).
\end{acknowledgments}

\newpage

\end{document}